# Broken keys to the kingdom

## Security and privacy aspects of RFID-based car keys

Jos Wetzels

a.l.g.m.wetzels@student.tue.nl

**Abstract.** This paper presents an overview of the current state-of-the-art of security and privacy concerns regarding RFID-based car key applications. We will first present a general overview of the technology and its evolution before moving on to an overview and discussion of the various known security weaknesses and attacks against such systems and the associated privacy risks they introduce.

**Keywords:** RFID, Car keys, Immobilizers, Keyless Entry, Security, Privacy

## 1   Introduction

Radio Frequency Identification (RFID) technology [1] has been around since 1973 in its modern form (with earlier incarnations going back all the way to Soviet espionage technology in 1945) but has only relatively recently been adopted on a mass-scale as the result of the convergence of low cost and increased sophistication. Security-oriented applications are myriad and include RFID-enabled car keys [2]. This paper will provide an overview of the security and privacy aspects of RFID-technology applied to car keys and the current state-of-the-art in these fields. With the introduction of Near-Field Communication (NFC) technology, which builds upon RFID, various proposals for smartphone-based NFC car keys have surfaced as well [3] but remain outside of the scope of this paper.

Car keys have multiple uses, from opening car doors, glove compartments and the trunk to starting the ignition. Traditional car keys are mechanical in nature and as such are vulnerable to the usual security risks: theft, key duplication, lock picking, etc [4]. In addition, cars can be hot-wired [5], a process which involves bypassing the car's ignition interlock in order to start the car without access to a key. RFID-technology has both sought to combat such traditional security issues as well as introduced new ones.

With user-convenience in mind, so-called remote keyless systems (RKS) [6] were first introduced for automobiles with the Renault Fuego in 1982 [7]. RKS refers to any lock that uses an electronic remote control as a key. Most RKS work through use of short-range radio transmissions, usually within a effective area of 5-20 meters. The functionality of RKS are usually present on a so-called key fob [8] and can be activat-ed when within the vicinity of the corresponding vehicle, the so-called "bubble", either by manually pressing the corresponding lock/unlock or automatically when pulling the door handle (in the case of so-called 'passive entry' systems). RKS includes both remote keyless entry systems (RKE), which refers to the functionality controlling the locking and unlocking of vehicle doors, as well as remote keyless ignition systems (RKI), which refers to starting the engine.

However, early RKS were not designed as an anti-theft measure but as a convenience measure [9], often implemented no security measures and did not mitigate problems such as lock picking or hot-wiring. In order to combat such security risks, ignition immobilizers [10] were developed. Almost all cars manufactured since 1995 have been outfitted with RFID transponder-enabled immobilizers [11] which have been mandatory in several countries since at least 1998 as well as all European insurance companies requiring them. Immobilizers are electronic security devices that prevent the vehicle engine from running unless the corresponding token is present. The setup works by having an RFID transponder present in the key fob of the car key (with later systems dropping the mechanical part of the key altogether) and an RFID transceiver (typically located close to the mechanical key lock in the steering column) connected to the vehicle's engine control unit (ECU) [12]. The transponder usually contains an EEPROM for storing additional configuration or secret data. Upon insertion of the ignition key in the ignition lock cylinder, the ECU will activate the transceiver to send a signal to activate the transponder (followed by a command). The transponder will responded with its coded message which the ECU will verify. Transponders are often passive but can be active as well, possessing an internal battery and behaving similar to passive devices except for increased performance and utilizing the battery to generate often more computationally complex response data [13]. In both cases, activation is mostly triggered by the transceiver signal. Upon authentication, the ECU will allow ignition to take place and fuel to flow. In the absence of such authentication (whether due the wrong key or no key being present at all), the engine will not start. In some modern vehicles, unauthorized starting attempts will initialize lock-outs, communicate an alert to a security firm by satellite or mobile phone or record information in the ECU related to speed, driver weight, geographic location, etc. for use by insurance investigations or warranty claims [10]. Many manufacturers have been moving away from mechanical keys altogether [14] in favor of fully electronic "smart keys" which incorporate all RKS functionality (including immobilizer disengagement) without the user having to insert a mechanical key in the ignition lock, provided they start ignition when inside the car. In initial implementations, RKE and immobilizer functionality was totally separated [9] but today both functions are combined in a single IC using a low frequency ferrite coil antenna for the transponder function and an UHF transmitter for the RKE function. In such scenarios, immobilizers are left as the only anti-theft mechanism.

## 2 Security

RFID-based car key systems have often been popularly portrayed as 'uncrackable' and are credited with a major reduction in car theft since their widespread adoption in the '90s [15,16]. However, in recent years a number of successful attacks have been developed against various RKE and immobilizer security schemes, which has led to some media attention [17,18,19,20] regarding the (in)security of RFID car key systems. In addition, there have been reports of criminals exploiting weaknesses in such systems [21,22,23], either in order to facilitate car theft or to obtain valuables locked inside the car. The security of RFID-based car key systems can roughly be divided in two aspects: (a) *Remote Keyless Entry systems* (b) *Immobilizers*. Sometimes, due to design restrictions, both systems will have shared security mechanisms and often there will be some overlap between utilized protocols and cryptographic algorithms.

### 2.1 RKE systems

As [24] shows there are several approaches to message transmission between vehicle and key fob, with varying levels of security:

1. *Fixed code*: Initially, RKE systems relied on fixed codes where a predetermined code is stored in the fob. Upon user activity, the fob transmits this fixed code and when the associated vehicle is within range it checks whether the code is valid. If this is the case, the associated predefined operation is performed.

2. *Rolling code*: This technique, also known as 'hopping code', utilizes a different code for each transmission and has been widely used in RKE systems. Rolling code systems maintain a sequence counter containing the code to be transmitted upon next activation of the fob. The transmitted sequence counter is usually encrypted (where key fob and vehicle share a secret key) using a block cipher, with KeeLoq being one of the most widespread ciphers used in this capacity [25]. After transmission, the sequence counter content is incremented. The vehicle also maintains a sequence counter. Upon reception of the fob's signal the receiver decrypts the sender's signal to obtain its sequence counter, which it then compares against its own sequence counter. The sender's message is validated and the operation is performed when the difference between the two sequence counters lies within a certain range. After receiving a valid code, the receiver also increments its sequence counter.

3. *Challenge-Response*: Authentication based on challenge-response schemes is widely used in immobilizer systems [14,26,27] but has found adoption in some RKE systems as well [28]. In a challenge-response scheme a bidirectional communication link is used and both the key fob and the vehicle share a secret encryption key. Upon activation (when the user pulls one of the door handles of the vehicle in a passive RKE scenario, for example) the vehicle sends a random number (the *challenge*) to the key fob. The key fob then encrypts the challenge using their shared encryption key and sends this *response* to the vehicle. The vehicle then verifies whether its encrypted challenge matches the response from the vehicle and when both match, the required operation is performed.

There are a number of attack scenarios possible [24] against the RKE approaches mentioned above. Many of the described attack scenarios introduce additional problems for victims regarding legal and insurance matters, as most of them leave little or no traces and proving actual theft has occurred (rather than fraud) is very hard.

1. *Brute-force attack*: Depending on the key size, RKE systems could be susceptible to straightforward brute-force attacks. In the case of a fixed code scheme an exhaustive search of the key space could be performed provided the key size is sufficiently small. In the case of a challenge-response scheme, a specific type of brute-force attack could be performed (known as a scanning attack) where upon every challenge, the same constant response is returned until a match occurs. Generally speaking, any authentication system with a big enough key space (or set of challenge-response pairs) should make brute-force attacks infeasible [29].

2. *Replay attack*: A replay attack, also known as a *'code-grabbing attack'*, involves an attacker intercepting the transmitted message from the target's key fob to the vehicle and storing it for later usage. In the case of a fixed code scheme, the attacker could easily re-transmit the captured message to automatically gain access to the vehicle since the message doesn't change each time the system is used. Rolling code schemes were developed to mitigate this kind of attack and, when properly implemented, provide adequate defense against replay attacks, as do challenge-response schemes. Since the sequence counter increments upon each successful action (in the case of a rolling code) and a new challenge is used for each authentication attempt (in the case of a challenge-response), simply capturing a valid message and retransmitting it later will no longer work. However, it is possible to successfully combine replay attacks with jamming attacks against some systems, as discussed below.

3. *Man-in-the-middle* (MITM) *attack*: A MITM-attack [24, 29, 30], also known as a *'two-thief attack'*, involves an attacker setting up an electronic 'bridge' between the target vehicle and the key fob. The attacker pulls a door handle of the vehicle (note that this assumes a scenario with passive RKE support) which initiates the transmission of a message which is captured by the attacker and sent to the other side of the 'bridge', which sends the message to the key fob. The key fob's response to this message is captured and sent back to the attacker near the vehicle, who then transmits it to the vehicle granting them access. Such a scenario is easiest performed with two attackers, but it isn't inconceivable passive equipment might be installed near a location the target frequents allowing for a single attacker scenario. Both rolling code and challenge response schemes are susceptible to MITM attacks. In addition, once an attacker has gained access to the vehicle by performing a relay attack on the RKE system, a subsequent relay attack could be performed against the key to disarm the ignition immobilizer (especially in the case of 'smart keys'), allowing the attacker to start the car and drive away. Since it would be very unsafe to turn off the engine on a running car once the key is absent, the relay doesn't have to be maintained (though some vehicles reportedly limit speed or send out a beeping signal in such a case [30]).

In order to mitigate against a MITM attack, several measures could be taken. A fairly simple one would involve shielding the key from RF signals while it isn't in use and commercial solutions for this type of protection are already available [31]. A more thorough form of protection, however, would involve a distance bounding scheme (taking into account the various attacks against such a scheme, like mafia- or terrorist fraud) which establishes a maximum upper bound on the physical distance between the vehicle and the key fob [32].

4. *Forward-prediction attack*: This type of attack involves the attacker attempting to predict a subsequent challenge by observing a set of previous challenges which can be obtained either by pulling the door handle several times (in the case of a passive system) or through observing a target entering and exiting their car a few times. If an attacker is capable of predicting such a subsequent challenge (because of an insecure Pseudo-Random Number Generator, for example), they can send out this challenge to the target's key fob while they're away from the vehicle, record the response and use it to open the vehicle. Rolling code schemes aren't vulnerable to such an attack since (assuming messages are encrypted and the key remains secret) it is impossible for an attacker to know what sequence counter an encrypted message corresponds to and as such even the ability to predict the subsequent counter wouldn't work.

In order to mitigate forward-prediction attacks, any challenge-response scheme should rely on a Cryptographically-Secure Pseudo-Random Number Generator (CSPRNG) for the generation of its challenges. In addition, the system should be designed such that challenge generation would halt for a fixed period of time if no valid responses have been received for a certain number of the previous challenges, thus decreasing the number of challenges generated per unit time and complicating an attacker's efforts further.

5. *Dictionary attack*: A dictionary attack involves an attacker building an electronic dictionary consisting of valid challenge-response keypairs. This is done by generating a random challenge while in this proximity of the key fob, transmitting it and collecting the response. Since there is no sender verification by the key fob, it will automatically respond. After building a sufficiently large dictionary, the attacker could go to the vehicle and keep re-initiating authentication (through pulling the door handle, for example) until a challenge is sent which is present in the dictionary, after which the corresponding response can be transmitted to unlock the vehicle.

In order to mitigate dictionary attacks, the key fob should halt challenge generation in the absence of valid responses as described above. Additionally, some form of sender verification should take place in the key fob. One such scenario [24] involves the vehicle adding an ID to every challenge, one part of the ID transmitted in plaintext and the other concatenated with the challenge and encrypted. The key fob would then compare the plaintext ID against a copy in its own memory and, if it's a match, decrypt the rest of the message. If the second part of the ID in the decrypted message matches the one in key fob memory it would re-encrypt the challenge using a different shared key and send it back to the vehicle as the response.

6. *Jamming attack*: A relatively well-known attack which has been utilized in real-world thefts [33] involves an attacker using a radio jammer. In the simplest scenario, an attacker would send out a (stronger) jamming signal at the same frequency as the targeted key fob. As such, when the target instructs their key fob to lock their car, the signal isn't received by the vehicle and as such it isn't locked. An attack like this, however, requires the target to pay no attention to the common visual (lights flashing) and auditory (horn beeping) confirmation signals most modern cars send out when being locked. A more sophisticated attack [34] would involve combining a jamming attack with a replay attack to utilize against rolling code schemes. In such a scenario, an attacker would jam the first attempt of the target to close their doors while intercepting the transmitted code. A second attempt to close the doors would be jammed as well, with the second code being intercepted again. Immediately after jamming the second closing attempt, the first intercepted code would be replayed against the vehicle, thus closing the doors and ensuring visual and auditory confirmation to the target. When the target leaves, the attacker can then use the intercepted second code in combination with a modified message to open the doors. Such an attack is rather delicate though since a single valid rolling code 'getting through' would invalidate all previously captured 'queuing' codes.

One way to mitigate such combined jamming and relay attacks would be to properly implement cryptographic nonces. Since such a nonce is used only once, the captured session and its associated nonce wouldn't correspond to the current session/nonce pair anymore upon the attacker's attempt to open the car. [35]

7. *Cryptographic and side-channel attacks:* In addition to the aforementioned attacks, it is also possible for an attacker to exploit weaknesses in the cryptographic algorithms (and their implementation) used in the communication protocols. Since protocol specifications and implementation details vary from vendor to vendor it is impossible to provide a generic oversight of all attacks of this type and as such we will focus on *KeeLoq*[36]. Developed in the mid '80s, it used to be the most commonly used algorithm for use in RKE systems and supports both rolling code (for RKE) and challenge-response (for immobilizer) modes. KeeLoq provides a straightforward rolling code scheme, with the addition of including a static shared *discriminant* value and *function* value specifying the task to be executed (open, lock, etc.) in each transmitted code as well as the sequence counter. The discriminant serves to identify the transponder to the receiver upon decryption. The KeeLoq cipher is a Non-linear Feedback Shift Register-based block cipher with a 32-bit block size, 64-bit key size and 528 rounds. The Original Equipment Manufacturer gets assigned a manufacturer key $k_m$ which is stored in all its receivers. When a new transponder is created, a key derivation function $f$ is called where unique device key $k_{dev} = f(ID, k_m)$. Derivation can be as simple as a XOR function or decryption using key $k_m$ to ensure irreversibility, even when $k_{dev}$ and ID are known. In addition, some implementations combine the ID with a random shared (32, 48 or 60 bit) seed before derivation. Receivers learn transponder keys by transponder transmission of the ID (keys are never sent in clear) after which the receiver performs key derivation and stores ID and $k_{dev}$. Since KeeLoq's widespread adoption for RKE purposes between 1995 and 2006 [37] various cryptographic attacks have been published [38, 39, 40, 41, 42], with various levels of practical applicability:

| Attack Type | Complexity | | |
|---|---|---|---|
| | Data | Time | Memory |
| Time-Memory Trade-Off | 2 CP | $2^{42.7}$ | $\approx 100$ TB |
| Slide/Algebraic | $2^{16}$ KP | $2^{65.4}$ | ? |
| Slide/Algebraic | $2^{16}$ KP | $2^{51.4}$ | ? |
| Slide/Guess-and-Determine | $2^{32}$ KP | $2^{52}$ | 16 GB |
| Slide/Guess-and-Determine | $2^{32}$ KP | $2^{50.6}$ | 16 GB |
| Slide/Cycle Structure | $2^{32}$ KP | $2^{39.4}$ | 16.5 GB |
| Slide/Cycle/Guess-and-Det.[a] | $2^{32}$ KP | $(2^{37})$ | 16.5 GB |
| Slide/Fixed Points | $2^{32}$ KP | $2^{27}$ | > 16 GB |
| Slide/Meet-in-the-Middle | $2^{16}$ KP | $2^{45.0}$ | $\approx 2$ MB |
| Slide/Meet-in-the-Middle | $2^{16}$ KP | $2^{44.5}$ | $\approx 3$ MB |
| Slide/Meet-in-the-Middle | $2^{16}$ CP | $2^{44.5}$ | $\approx 2$ MB |
| Time-Memory-Data Trade-Off | 68 CP, 34 RK | $2^{39.3}$ | $\approx 10$ TB |
| Related Key | 66 CP, 34 RK$^\gg$ | negligible | negligible |
| Related Key | 512 CP, 2 RK$^\gg$ | $2^{32}$ | negligible |
| Related Key/Slide/MitM | $2^{17}$ CP, 2 RK$^\oplus$ | $2^{41.9}$ | $\approx 16$ MB |

**Table 1.** KeeLoq cryptanalytical attack overview [28].

The best cryptanalytical attack to date [28] is a combined *slide* and *meet-in-the-middle* attack possible largely due to the fact that KeeLoq's cipher consists solemnly of a multi-round application of the same function $F$ weak against a known-plaintext attack. The attack targets the challenge-response mode of KeeLoq (also referred to as *Identify-Friend-or-Foe* mode) and requires $2^{16}$ known plaintexts from one transponder (requiring 65 minutes to obtain this data) and 7.8 days of calculations on 64 CPU cores. A variant of the attack, requiring $2^{16}$ chosen plaintexts, takes 3.4 days of calculations on 64 CPU cores. This attack targets the challenge-response mode (and not the *rolling code* mode, which was the dominant commercial application) since challenges are not authenticated and as such an attacker can obtain as many chosen plaintexts as necessary within a reasonable timeframe, while in rolling code mode the attacker only has access to the ciphertexts. However, this only allows for simple transponder cloning. Only in the case where a weak key derivation method is used is it possible to obtain the manufacturer key required to generate keys for a new transponder. While the aforementioned cryptanalytic attacks show the inherent weakness of the KeeLoq design, they long proved impractical in real-world applications and were overshadowed by the fact that far faster brute force attacks (recovering keys derived from 48-bit seeds in under 6 hours) using dedicated hardware existed [43]. Mitigation of such cryptanalytic attacks would involve the adoption of strong cryptographic algorithms, such as AES [44].

However, with the discovery of the first differential power analysis (DPA) attack on KeeLoq [45], the situation changed. The DPA attack in question (targeting variations in power consumption related to the cipher's state register) presented a complete break of KeeLoq and enabled an attacker to extract both the manufacturer key present in the receiver (which implements the cipher in software) and the device key present in the transponder (implementing the cipher in hardware) by performing DPA on captured power traces. In contrast to the cryptanalytic attacks discussed above, this side-channel attack is applicable to both challenge-response and rolling code modes. The practical results of the paper meant that an attacker, requiring only two intercepted messages, could clone any transponder in a manner of minutes (from 5-30 power traces) and allowed the attacker with access to a receiver to recover the manufacturer key and generate new transponders at will. Recovering the manufacturer key from the receiver, however, proved to be more complicated and time consuming (requiring ~10000 traces, taking hours), meaning that attackers would potentially have to outsource this part to criminal cryptographers. The attack works for all key derivation schemes, instantly in the case of a weak scheme or within reasonable time using dedicated hardware in the case of a stronger scheme. While the authors of the paper initially suggest a sufficiently random 60-bit seed as mitigation, research into cost-optimized dedicated hardware [46] has shown that even such an implementation could be broken within a reasonable amount of time given adequate investment (as the attack is fully parallelizable). Further improvements upon the attack [47] allowed an attacker to extract the manufacturer key with a single measurement using Simple Power Analysis (SPA), thus greatly reducing the complexity and increasing the efficiency of the attack. The SPA attack was possible due to the software implementation present in the receiver consuming a constant amount of clock cycles for every operation except the NLF lookup table, making execution time vary for different cipher texts and leaving it vulnerable to SPA. It is noted that while countermeasures might be undertaken to mitigate SPA and DPA (such as constant run times), other types of physical attacks such as fault injection might pose another future threat in addition to the fact that the current non-constant run times also render the implementation potentially vulnerable to timing attacks.

## 2.2 Immobilizers

Immobilizer systems almost always use a challenge-response protocol for message transmission and authentication between vehicle and key fob. Since the difference between immobilizer security systems and RKE systems is mostly a question of implementation (in that transmission distance is far shorter, usually requires physical key presence in the mechanical lock and operates in challenge-response mode), some of the security mechanisms and associated attacks that apply to RKE systems apply to immobilizers as well, albeit with certain variations. For the sake of brevity we will not reiterate the attack categories described in 2.1, as they apply to immobilizer security as well, but merely highlight the differences before discussing attacks specific to immobilizer security systems. A key difference separating attacks targeting immobilizer systems from attacks aimed at RKE systems lies in the fact that immobilizer communication only occurs within a certain small 'safe' radius around the immobilizer receiver. This radius is usually limited to a few centimeters in the case of immobilizers that require mechanical key presence before activation and a few meters (generally the 'inside' of the car) in case of 'smart key' immobilizers. As such, a prerequisite for attacks on immobilizer systems is access to the car interior, requiring an attacker to force their way in either physically (eg. smashing a window) or bypassing RKE security. This difference does not, however, make MITM- or dictionary attacks any less feasible as construction of the dictionary can still take place (though they are rather inefficient for most implementations [48]) with close proximity to the key fob (which responds to the challenges initiated by the vehicle) and both attacks can be executed given that entry to the vehicle is obtained. In addition, certain immobilizer implementations [49] allow for retransmission of the last response the vehicle sent out, allowing the attacker to use the extra time for the repeated message to get the original response and relay it to the vehicle, hence circumventing distance-bounding countermeasures. Generally speaking all scenarios that apply to challenge-response protocols mentioned in 2.1 apply to immobilizer systems as well, sometimes under condition of proximity, except for jamming attacks as those were aimed specifically at preventing the closing of car doors.

One real-world type of attack [50,51,52], specifically aimed at immobilizer systems and largely targeting BMWs, involved attackers obtaining access to the vehicle (usually through a crude jamming attack) and plugging in a diagnostic tool (produced for garages and widely available on the market) into the vehicle's On-Board Diagnostics (OBD) port to retrieve the immobilizer key so that a cloned transponder could be fabricated later on. A covert GPS tracker would be installed so the car could be stolen with the cloned transponder at an opportune moment. Such attacks are possible because access to the OBD port is unauthenticated and because manufacturers rarely secure access to the OBD port [53], which has spawned various after-market solutions aimed at providing some form of access control to the OBD port, such as keypad-protected covers. In addition to facilitating car theft through allowing attackers to bypass immobilizer systems, the plethora of security problems surrounding OBD [54] have opened up a pandora's box [55] of additional threats which fall outside the scope of this paper.

In order to illustrate the problems surrounding immobilizer security, we will proceed to discuss the publicized developments in attacks on the three technologies that dominate the immobilizer market [56]:

1. *DST:* Texas Instruments' Digital Signature Transponder (DST) [48] is a cryptographically enabled RFID transponder developed in 1995 and is used for immobilizer authentication purposes as well as (formerly) for the Exxon-Mobil Speedpass payment system. The DST scheme follows the usual immobilizer setup with a passive transponder and a challenge-response protocol using a proprietary block cipher. The first version of DST that was rolled out, known as DST-40, is a block cipher that used a 40-bit key to produce a 40-bit ciphertext which was truncated to a 24-bit response to the 40-bit challenge supplied by the vehicle. In 2005, a group of researchers at the Johns Hopkins University [26] reverse-engineered the until-then undisclosed cipher, performed a brute-force attack and managed to recover the key for use in transponder cloning. The researchers programmed a series of FPGAs to execute a brute-force attack on the basis of known challenge-response pairs, recovering a key from two pairs in roughly 11 hours using a single FPGA and in less than an hour when using 16 FPGAs. In addition, they used an FPGA to compute a series of pre-computed look-up tables reducing the cracking time to seconds.

   In order to clone a transponder an attacker would have to obtain two challenge-response pairs from the target transponder in order to derive the key. One scenario would involve an active attack where the attacker sends out a chosen challenge to the target transponder and capture the response. Since a DST responds to as many as eight queries per second this phase can be accomplished fast enough to ensure that even brief physical proximity (a prerequisite in the case of immobilizer systems) of the attacker to the target and their transponder key is enough. Since the attacker can choose their challenge, they can pre-compute the associated look-up tables to drastically reduce the time required to recover the key. This would allow an attacker to create a small self-contained cloning device, containing pre-computed lookup tables, which automatically clones any transponder key near to it upon activation. Another scenario would involve a passive attack where an attacker uses eaves-dropping equipment (with a reported maximum distance of a few meters) to intercept legitimate communications between vehicle and transponder and perform an FPGA-supported brute-force attack later on. This weakness is due to the short key length of the DST-40 cipher though it is claimed that cryptanalytic attacks against the slightly (but still insufficiently) stronger variant DST-80 are possible as well [57], although no sources could be found to confirm that claim. In order to mitigate such attacks on DSTs (short of recalling all transponders and increasing the key length) it is possible to use RFID shielding pouches or wallets [31] or use a so-called RSA blocker tag [58], which responds positively to all unauthorized requests, hence interfering with the attacker's communication with the target transponder.

2. *Hitag*: Philips/NXP's Hitag [14] has several variants used primarily in RFID transponder systems for immobilizer and RKE purposes (especially as a backup mechanism upon smart key power depletion). Introduced in 1996, the Hitag2 variant of Hitag is one of the most widely used transponder systems in the world. Hitag supports three modes of operations:

   (i) *Public*: Where the contents of the transponder memory are simply broadcast in plaintext upon power up.

   (ii) *Password*: Receiver and transponder authenticate each other by exchanging their passwords which, however, is transmitted in the clear and hence vulnerable to replay attacks.

   (iii) *Crypto*: In this mode, which is the one used for immobilizer systems, Hitag receiver and transponder share a common secret key. At the start of each session, the vehicle's receiver sends a command to the transponder which replies with its serial number. To mitigate replay attacks, an Initialization Vector (IV) is generated for each session using the secret key, IV and tag serial number for initialization of the proprietary Hitag stream cipher. After initialization Hitag produces a keystream where the first 32-bits are used as an authenticator and the remaining 32-bits are used for encryption. The receiver then sends the 32-bit IV and 32-bit authenticator to the transponder. The transponder calculates the authenticator as well and, if they match, responds by encrypting (using a proprietary stream cipher) its configuration data and password (used as a fallback mechanism in case the cryptosystem is broken) using the rest of the key and sends this to the receiver. The difference between Hitag1 and Hitag2 lies mainly in the fact that Hitag2 uses a 48-bit secret key as opposed to Hitag1's 32-bit. Since the receiver is authenticated first, this prevents tag-only attacks (eg. replay attacks or standard dictionary attacks) and requires an attacker to obtain sniffed data first.

Since Hitag1 relies on a mere 32-bit secret key, a brute-force attack is reported to be feasible within the order of mere minutes [56,57]. Research [59] has also shown that Hitag1 and Hitag2 are vulnerable to modified dictionary attacks. Since Hitag supports authenticated challenge-response sessions, a so-called Three-step dictionary attack first gathers a series of challenges with a valid authenticator by triggering the receiver and before initiating a regular dictionary attack. Hitag's vulnerability to such an attack could be mitigated by implementing additional countermeasures. For example, the number of successive unsuccessful authentication sessions on the receiver could be limited (with a timeout for example) as well as limiting the number of authentication sessions on the transponder side. A more general mitigation would involve improving the size of the challenge-response pair space or the amount of time it takes for a session to complete in order to increase the complexity of dictionary attacks to the point of infeasibility.

With the market prominence of Hitag2, follow-up attacks have been developed targeting various aspects of its design and implementation:

| Attack | Description | Practical | Computation | Traces | Time |
|---|---|---|---|---|---|
| [64] | Brute-force | Yes | 2102400 min | 2 | 4 years |
| [60] | Sat-solver | Yes | 2880 min | 4 | 2 days |
| [65] | Brute-force | Yes | 660 min | 2 | 11 hrs |
| [62] | Sat-solver | No | 386 min | N/A | N/A |
| [64] | Brute-force | Yes | 103.5 min | 2 | 2 hrs |
| [14] | Cryptanalytic | Yes | 5 min | 136 | 6 min |
| [63] | Cube | No | 1 min | 500 | N/A |

**Table 2.** Attacks against Hitag2 (partially based on [14])

Initial research [60] focused on cryptanalytic attacks against the Hitag2 cipher which showed that reducing the cipher to a series of multivariate quadratic equations and solving it using a SAT-solver [61] is possible given that the algorithm is very linear (eg. output isn't filtered back into the state) and doesn't have too many rounds. The attack uses the fact that 32 bits of the key stream are known per IV to mount a known IV attack that takes a little under 2 days on a full 48-bit key (a chosen IV attack taking only 6 hours is possible against the cipher in general but not against its application in the described protocol. Similarly, a SAT-solver optimization was described [62] making an attack taking under 7 hours possible which, however, required 50 bits of contiguous key stream making it theoretical). The second cryptanalytic attack developed against Hitag2 concerned a so-called cube attack [63] taking only a minute on your average PC. However, since cube attacks are chosen IV attacks, the actual protocol implementation of Hitag2 makes this attack strictly theoretical. In order to mitigate such attacks, cipher security (as well as key size) should be increased, preferably by using publicly reviewed strong cryptography.

In addition to the aforementioned cryptanalytic attacks, Hitag2 has also been the target of brute-force attacks. Such an attack [64] involved sniffing a single session between vehicle and transponder to obtain the transponder serial number, the IV and the authenticator before generating and testing all possible $2^{48}$ keys. Every key is passed through Hitag2 and the resulting authenticator is compared to the captured authenticator, a match resulting in the key becoming a candidate. In or-der to find the correct candidate, all candidates are checked against a second sniffed session (which has a different IV and authenticator). The matching candidate is the recovered secret key. Brute-force performance on a regular Pentium-IV processor would take 4 years and as such is impractical but a dedicated FPGA-based platform (a so-called COPACOBANA) yielded far better results, recovering the key in less than two hours. Since such hardware is fairly expensive, an attack running on off-the-shelf hardware [65] using OpenCL across CPUs and GPUs was developed capable of key recover in less than 11 hours on a budget of €5,000.

Finally, a paper published in 2012 [14] revealed a whole host of other weaknesses in and practical attacks against Hitag2. The core weaknesses revealed by the paper are as follows:

(i) *Arbitrary length key stream oracle*: It is possible for an attacker, without knowing the secret key and with only a single intercepted authentication attempt, to obtain an arbitrary length of key stream from the transponder. Hitag2 transponders offer the possibility to extend the length of any command with a multiple of five bits (up to a total message length of 5010 bits) using redundancy messages. Since the receiver is not authenticated to the transponder (the transponder lacks a PRNG), it is possible to replay any valid nonce & authenticator pair to the transponder. The transponder will only respond when the received message is decrypted correctly. By replaying the intercepted pair and gradually extending the length with redundancy messages (5 bits at a time), the attacker only has to brute-force 32-bits per replay to obtain the corresponding key stream.

(ii) *Dependencies between sessions*: At a specific state the cipher is fully initialized and from there on only produces key streams. This reveals that the 48-bit cipher internal state is randomized by a nonce of only 32 bits, as such, at that specific state, only the upper half of the LFSR are affected by the nonce, the lower half remaining constant, causing strong dependency throughout sessions as these 16 persistent bits correspond to bits $k_0..k_{15}$ of the secret key.

(iii) *Low degree determination of the filter function*: Due to a weakness in the cipher design (4 filtered bits cover 14 bits of the internal state) there's a chance of 1 in 4 that an output bit of the cipher is determined by only 34 out of 48 bits of the internal state. Consequentially $1/4^{th}$ (on average) of all authentication attempts leaks one bit of secret key information.

These weaknesses can be exploited in the following attack scenarios:

(i) *Malleability attack*: In this attack the lack of a PRNG or challenge to the receiver on part of the transponder is exploited by having an attacker inter-cept a valid authentication attempt. Exploiting the key stream oracle weakness, the attacker can read known plaintext and recover the key stream. Subsequently the attacker can then use the key stream to read any other memory block including the secret key. If properly configured, however, the secret key cannot be read by the transponder, though most implementations appear to be incorrectly configured.

(ii) *Time/memory tradeoff attack*: A second attack (slower but generally applicable to LSFR-based stream ciphers) relies on a time/memory tradeoff to generate a lookup table (of roughly 1.2TB, a process which takes less than a day) of $2^{37}$ cipher state/key stream pairs. The attacker then emulates a transponder to obtain a nonce-authenticator pair and replays it to the target transponder exploiting the key stream oracle to obtain 256 bytes of contiguous key stream. The attacker then iterates through the obtained key stream in blocks of 48-bits and looks them up in the table. If a match is found, it is recorded as a candidate internal state. The rest of the key stream is used to verify the internal state. Using the obtained internal state, the attacker performs a rollback of the cipher to obtain the secret key. Total attack time is roughly one minute (30 seconds to gather 256 bytes of key stream from the transponder and worst case 30 seconds to perform the lookup).

(iii) *Cryptanalytic attack*: The third and most powerful attack requires only 136 authentication attempts from the vehicle (gathered by the attacker emulating a transponder) in order to recover the secret key (given that the attackers knows a valid transponder serial number). The attacker guesses the first 34 key bits and performs a check on the authenticator (using the low determination flaw) to eliminate non-candidate keys. The session dependency flaw allows the attacker to perform this test many times (136/4 = 34 bits), increasing the efficiency of non-candidate key elimination. For the few candidate keys that pass these tests (~2-3) an exhaustive search for the remaining 14 bits is performed. The entire attack can be performed in 360 seconds.

In addition to the above design flaws, the paper also mentions that a lot of vendors make serious implementation mistakes, reducing security even further. The implementation errors encountered were: default or predictable transponder passwords, usage in keyless ignition systems which bypass the need for a physical key being inserted, weak PRNGs on the side of the vehicle (sometimes resulting in nonces with an entropy of as little as 8 bits), secret keys with repetitive patterns, poorly configured transponders (eg. without a set protection bit and thus readable secret keys). All tested cars did however have transponder serial number white listing and as such an attacker would have to obtain one from the target transponder first before executing an attack. In order to mitigate such attacks migration to a cryptographically secure cipher would be recommended as well as extending the transponder password and introducing communication delay after au-thentication failure. In addition, using CSPRNGs and introducing transponder authentication to the protocol would also improve overall security. NXP's Hitag AES solution already implements the (as of writing) cryptographically secure AES, though the associated protocol is proprietary and as such nothing can be said about the level of security it offers.

3. *Megamos*: EM's Megamos is another immobilizer system with a significant market share. Recently, research [27] seemed to indicate that Megamos was vulnerable to attacks comparable to those against Hitag2 with a similar level of practicality. Apparently[66], the researchers did not reverse engineer a Megamos immoblizer system itself but a piece of 3$^{rd}$ party software implementing a transponder programmer. The researchers managed to identify 3 weaknesses: weak secret keys (even though it is claimed [67] that the cipher used in Megamos relies on a 96-bit secret key), poor key updating and a cryptanalytical attack. However, due to a temporary judicial injunction, publication of both the paper and presentation have been prohibited and as such the nature of the weaknesses remains publicly unavailable. One presentation [56], however, superficially mentions at least one vulnerability in the Megamos protocol (which was ironically described rather clearly in the court injunction itself) due to an insecure challenge-response protocol where the transponder doesn't authenticate itself to the vehicle.

## 2.3 The big picture

In conclusion, the security problems associated with both RKE and immobilizer systems come down to the following three points (though specifically discussing immobilizers, they're equally applicable to RKE systems) [57]:

(i) *Legacy*: Phasing out immobilizers seems to take years (eg. DST-40 was broken in 2005 and phased out only towards 2011). Even after certain systems have been broken, manufacturers have continued to use them for long periods of time, something which shouldn't be the case.

(ii) *Proprietary and weak cryptography*: All discussed immobilizer systems use proprietary cryptographic algorithms which are often very weak and have gone without public scrutiny or serious cryptanalysis before being rolled out. Memory and power consumption limitations on RFID devices imposes certain limitations on cryptographic functionality but the existence of devices implementing strong cryptographic algorithms (such as AES) shows reliance upon weak and proprietary cryptography is unnecessary. While this tendency toward violating Kerckhoff's principle is widespread, something is to be said for not disclosing implemented algorithms since knowledge of the used algorithm eases the task of attackers looking for potential side channel-based weaknesses and mitigating side-channel attacks is costly. Security through obscurity doesn't work but obscurity in addition to security does complicate (especially in terms of investment of resources

(iii) *Implementation faults*: In addition to (ii) implementation faults further lower immobilizer security, often reducing the complexity of certain attacks or even undoing any security features altogether. Thorough application of 'best practices' and security analysis of developed or integrated systems ought to be standard practice.

and effort) an attacker's job. Nevertheless, migration to strong ciphers (whether proprietary or public) is eventually necessary.

Considering the spread of "car digitalization" and the expansion of the attack surface this presents [55], Karsten Nohl notes [56] this leaves manufacturers with two options: a) keep cars simple and hence the attack surface limited and controlled b) strongly invest in serious security to eliminate design and implementation flaws, while phasing out broken systems fast.

## 3  Privacy

In stark contrast to security-related issues there has been almost no public discussion about or research regarding the privacy aspects of RFID-based car keys. While there has been plenty of research regarding RFID privacy issues in general [68,69,73] and some discussion about RFID privacy issues surrounding cars in particular [70], neither RKE or immobilizer applications seem to be of much privacy-related interest. As such, this lack of public attention towards and research regarding the privacy aspects of RFID car keys means this overview (which does not include original research) will be limited in this regard. Apart from the obvious risks to general privacy posed by the insecurity of RFID car key applications it is also worth mentioning that at least one aspect of such systems falls under the broad umbrella of 'privacy concerns' proper. Since virtually all transponders send out their unique serial number at some point during a communication session, this constitutes a piece of 'personal information'. California legislation [71], for example, defines personal information to be, among other things, "any unique personal identifier". Even though such serial numbers have no meaning by themselves, they derive their meaning from their relationship to the person carrying them and their uniqueness makes it possible to both couple a vehicle and its related information to a person, as well as track said person's movement patterns as Bruce Schneier [72] remarks. While privacy risks are somewhat reduced by the fact that transponders operate at relatively short ranges, a scenario involving targeted individual attacks (eg. following or approaching someone closely in order to query their transponder for information) or 'turnstile-style' funneling of personal movements combined with automated querying isn't far-fetched. Despite approaches suggested to mitigate privacy risks associated with RFID [73], car keys haven't been the subject of research into this area. In conclusion, the privacy aspects of RFID-enabled car keys seem woefully under-investigated and would benefit from research in that direction. Judging from the little information that is available, relying on RF shielded pouches seems to be the most reliable privacy measure overall.

# 4    Conclusion

As we can see, RFID-based car key applications (whether as RKE or immobilizer systems) have historically shown a woeful lack of adequate security levels, being vulnerable to a plethora of attacks resulting from a mix of flawed protocol designs and the usage of weak cryptography, with often a host of additional implementation faults on top. And while recently RFID car key implementations with a more serious level of security have been surfacing on the market so far these don't seem to have been rolled out on a grand scale. Considering the relatively low cost of manufacturing (after initial research investment) car key systems with a more adequate level of security (given advances in implementing strong cryptography for RFID applications) compared to the price of certain high-end car models, it seems paramount that the automotive industry take steps to ensure vehicle security for their consumers. In this respect, it is imperative to deal with the three issues mentioned in section 2.3: a) legacy problems b) weak and proprietary cryptography c) implementation faults.

Regarding matters of privacy, it is recommended that research be undertaken to investigate the privacy risks associated with RFID-based car key applications to fill the current gap in that area. In the meantime, the best option for concerned car owners seems to be the use of shielded pouches.

**References**

1. http://en.wikipedia.org/wiki/Radio-frequency_identification. Last visited 16-03-2014
2. Ron Weinstein. *RFID: A Technical Overview and Its Application to the Enterprise*. In IT Pro, 01-05-2005
3. Christoph Busold et al. *Smart Keys for Cyber-Cars: Secure Smartphone-based NFC-enabled Car Immobilizer*. In CODASPY'13, 18-02-2013
4. http://en.wikipedia.org/wiki/Motor_vehicle_theft#Methods. Last visited 16-03-2014
5. http://en.wikipedia.org/wiki/Hot-wiring. Last visited 16-03-2014
6. http://en.wikipedia.org/wiki/Remote_keyless_system. Last visited 16-03-2014
7. http://www.classicandperformancecar.com/front_website/octane_interact/carspecs.php/?see=3341. Last visited 16-03-2014
8. http://searchsecurity.techtarget.com/definition/key-fob. Last visited 16-03-2014
9. Paris Kitsos et al. *RFID Security: Techniques, Protocols and System-On-Chip Design*. 2008
10. http://en.wikipedia.org/wiki/Immobilizer. Last visited 16-03-2014
11. http://www.rfidjournal.com/blogs/experts/entry?8162. Last visited 16-03-2014
12. http://en.wikipedia.org/wiki/Engine_control_unit. Last visited 16-03-2014
13. Andreas Hagl et al. *RFID: Fundamentals and Applications. In RFID Security: Techniques, Protocols and System-On-Chip Design*, 2008
14. Roel Verdult et al. *Gone in 360 seconds: Hijacking with Hitag2*. In Security'12 Proceedings of the 21st USENIX conference on Security symposium, 2012
15. http://online.wsj.com/news/articles/SB112811826476357203. Last visited 16-03-2014
16. http://www.rfidjournal.com/blogs/experts/entry?8162. Last visited 16-03-2014
17. http://www.wired.com/wired/archive/14.08/carkey.html. Last visited 16-03-2014
18. http://www.nytimes.com/2005/01/28/science/28cnd-key.html?_r=0. Last visited 16-03-2014
19. http://www.engadget.com/2011/01/16/research-shocker-keyless-car-entry-systems-can-be-hacked-easily/. Last visited 16-03-2014
20. http://www.motorauthority.com/news/1024586_scientists-crack-keyless-entry-security-systems. Last visited 16-03-2014
21. http://hackaday.com/2012/07/07/keyless-bmw-cars-prove-to-be-very-easy-to-steal/. Last visited 16-03-2014
22. http://www.today.com/news/police-admit-theyre-stumped-mystery-car-thefts-6C10169993. Last visited 16-03-2014
23. http://www.telegraph.co.uk/news/uknews/crime/9623150/Lock-jammers-steal-cars-in-a-click.html. Last visited 16-03-2014
24. Ansaf Ibrahem Alrabady et al. *Analysis of Attacks Against the Security of Keyless-Entry Systems for Vehicles and Suggestions for Improved Designs*. In IEEE TRANSACTIONS ON VEHICULAR TECHNOLOGY, VOL. 54, NO. 1, JANUARY 2005
25. Kobus Marneweck. *An Introduction to Keeloq Code Hopping*. 1996
26. Stephen C. Bono et al. *Security Analysis of a Cryptographically-Enabled RFID Device*. In Proceedings of the 14th conference on USENIX Security Symposium - Volume 14, 2005
27. Roel Verdult et al. *Dismantling Megamos Crypto: Wirelessly Lockpicking a Vehicle Immobilizer*. In 22nd USENIX Security Symposium (USENIX Security 2013). USENIX Association, 2013
28. Sebastiaan Indesteege et al. *A Practical Attack on KeeLoq*. In Lecture Notes in Computer Science Volume 4965, 2008
29. Ansaf Ibrahem Alrabady et al. *Some Attacks Against Vehicles' Passive Entry Security Systems and Their Solutions*. In IEEE TRANSACTIONS ON VEHICULAR TECHNOLOGY, VOL. 52, NO. 2, MARCH 2003

30. Aurelien Francillon et al. *Relay Attacks on Passive Keyless Entry and Start Systems in Modern Cars*. In Proceedings of the Network and Distributed System Security Symposium, NDSS 2011
31. http://www.epiguard.ch/ki/en/RFID-car-key-protection-55.html. Last visited 16-03-2014
32. Tao Yang et al. *Resisting Relay Attacks on Vehicular Passive Keyless Entry and Start Systems*. In 2012 9th International Conference on Fuzzy Systems and Knowledge Discovery (FSKD 2012), 2012
33. http://www.telegraph.co.uk/news/uknews/crime/9623150/Lock-jammers-steal-cars-in-a-click.html. Last visited 16-03-2014
34. Markus Kasper et al. *Breaking KeeLoq in a Flash: On Extracting Keys at Lightning Speed*. In Lecture Notes in Computer Science Volume 5580, 2009
35. http://en.wikipedia.org/wiki/Cryptographic_nonce. Last visited 16-03-2014
36. http://en.wikipedia.org/wiki/KeeLoq. Last visited 16-03-2014
37. Thomas Eisenbarth et al. *Messing around with Garage Doors: Breaking KeeLoq with Power Analysis*. In 25C3, 2008
38. Andrey Bogdanov. *Cryptanalysis of the KeeLoq block cipher*. In Cryptology ePrint Archive, Report 2007/055
39. Andrey Bogdanov. *Attacks on the KeeLoq Block Cipher and Authentication Systems*. In 3rd Conference on RFID Security, 2007
40. Nicolas T. Courtois et al. *Algebraic and Slide Attacks on KeeLoq*. In Cryptology ePrint Archive, Report 2007/062
41. Nicolas T. Courtois et al. *Algebraic and Slide Attacks on KeeLoq*. In Proceedings of Fast Software Encryption 2008
42. Thomas Eisenbarth et al. *Physical Cryptanalysis of KeeLoq Code Hopping Applications*. In Cryptology ePrint Archive, Report 2008/058
43. Tim Guneysu et al. *High-Performance Cryptanalysis on RIVYERA and COPACOBANA Computing Systems*. In High-Performance Computing Using FPGAs, 2013
44. Xiao Ni et al. *AES Security Protocol Implementation for Automobile Remote Keyless System*. In Vehicular Technology Conference, 2007. VTC2007-Spring. IEEE 65th, 2007
45. Thomas Eisenbarth et al. *On the Power of Power Analysis in the Real World: A Complete Break of the KeeLoq Code Hopping Scheme*. In Lecture Notes in Computer Science Volume 5157, 2008
46. Martin Novotny et al. *Cryptanalysis of KeeLoq with COPACOBANA*. In Special-purpose Hardware for Attacking Cryptographic Systems, 2009
47. Christof Paar et al. *KeeLoq and Side-Channel Analysis—Evolution of an Attack*. In 2009 Workshop on Fault Diagnosis and Tolerance in Cryptography (FDTC), 2009
48. Ulrich Kaiser. *Digital Signature Transponder*. In RFID Security: Techniques, Protocols and System-on-Chip Design, 2008
49. Stefan Tillich et al. *Security Analysis of an Open Car Immobilizer Protocol* Stack. In Lecture Notes in Computer Science Volume 7711, 2012
50. http://www.technewsdaily.com/7932-high-tech-car-theft.html. Last visited 16-03-2014
51. http://www.telegraph.co.uk/news/uknews/crime/9369783/Thieves-placed-bugs-and-hacked-onboard-computers-of-luxury-cars.html. Last visited 16-03-2014
52. http://www.theregister.co.uk/2012/09/17/bmw_car_theft_hack/. Last visited 16-03-2014
53. http://www.sbd.co.uk/obd-port-security-whos-got-it-covered/. Last visited 16-03-2014
54. Rob Vandenbrink. *Dude your car is pwnd*. In SANSFIRE, 2012
55. Stephen Checkoway et al. *Comprehensive Experimental Analyses of Automotive Attack Surfaces*. In SEC'11 Proceedings of the 20th USENIX conference on Security, 2011
56. Karsten Nohl. *Car immobilizer hacking*. In SIGINT13, 2013

57. Karsten Nohl. *Immobilizer security*. In ESCAR 2010
58. http://en.wikipedia.org/wiki/RSA_blocker_tag. Last visited 16-03-2014
59. Andrey Bogdanov et al. *On the Security and Efficiency of Real-World Lightweight Authentication Protocols*. In 1st Workshop on Secure Component and System Identification (SECSI 2008), 2008
60. Nicolas T. Courtois et al. *Practical Algebraic Attacks on the Hitag2 Stream Cipher*. In Lecture Notes in Computer Science Volume 5735, 2009
61. http://en.wikipedia.org/wiki/Boolean_satisfiability_problem. Last visited 16-03-2014
62. Mate Soos et al. *Extending SAT solvers to cryptographic problems*. In heory and Applications of Satisfiability Testing-SAT 2009
63. Siwei Sun et al. *Cube Cryptanalysis of Hitag2 Stream Cipher*. In CANS'11 Proceedings of the 10th international conference on Cryptology and Network Security, 2011
64. Petr Stembera et al. *Breaking Hitag2 with Reconfigurable Hardware*. In 14th Euromicro Conference on Digital System Design, 2011
65. Vincent Immler. *Breaking Hitag 2 Revisited*. In SPACE'12 Proceedings of the Second international conference on Security, Privacy, and Applied Cryptography Engineering, 2012
66. Robert Carolina et al. *Megamos Crypto, Responsible Disclosure, and the Chilling Effect of Volkswagen Aktiengesellschaft vs Garcia, et al*. In ISG Research Seminar 3 October 2013
67. http://www.bicotech.com/doc/megamos_cr.pdf. Last visited 16-03-2014
68. Robert H. Deng. *A New Framework for RFID Privacy*. In ACNS'12 Proceedings of the 10th international conference on Applied Cryptography and Network Security, 2012
69. Gianmarco Baldini et al. *RFID Tags Privacy Threats and Countermeasures*. In European Commission JRC Scientific and policy reports, 2012
70. http://www.sleeandtopher.com/how-rfid-chips-are-making-your-identity-harder-to-protect/. Last visited 16-03-2014
71. Erwin Schmidt et al. *RFID and Privacy*. In 4th European Workshop on RFID Systems and Technologies (RFID SysTech), 2008
72. https://www.schneier.com/blog/archives/2005/10/rfid_car_keys.html. Last visited 16-03-2014
73. Simson L. Garfinkel et al. *RFID Privacy: An Overview of Problems and Proposed Solutions*. In Security & Privacy, IEEE (Volume:3 , Issue: 3 ), 2005